\documentclass[pre,twocolumn,superscriptaddress,showpacs,floatfix,lengthcheck]{revtex4-1}
\usepackage{graphicx} \usepackage{amsmath} \usepackage{times}
\usepackage{amssymb} \usepackage{mathrsfs} \usepackage{mathenv}
\usepackage{chemarr} \usepackage{color}
\definecolor{linkcolor}{rgb}{0,0,0.6} \usepackage[
  pdftex,colorlinks=true, pdfstartview=FitV, linkcolor= linkcolor,
  citecolor= linkcolor, urlcolor= linkcolor, hyperindex=true,
  hyperfigures=false] {hyperref}
	
\usepackage{url}
\usepackage{version}

\providecommand{\addhyphen}[1]{#1.---}

\makeatletter
\renewcommand \paragraph{%
  \@startsection
    {paragraph}%
    {4}%
    {\parindent}%
    {\z@}%
    {-.1em}%
    {\normalfont\normalsize\itshape\addhyphen}%
}%
\makeatother

\newcommand{\cO}{\mathcal{O}}

\newcommand{\cA}{\mathcal{S}}
\newcommand{\cP}{\mathcal{P}}

\newcommand{\C}{\mathcal{C}}
\newcommand{\dd}{\text{d}}

\newcommand{\ee}{\text{e}}

\newcommand{\p}{\partial}

\newcommand{\br}{\text{\bf r}}

\newcommand{\eps}{\varepsilon}

\newcommand{\kb}{}

\newcommand{\bxi}{\boldsymbol{\xi}}

\newcommand{\bva}{{\bf v}_\text{\tiny A}}
\newcommand{\brr}{{\bf r}}

\newcommand{\va}{v_\text{\tiny A}}

\newcommand{\td}{\tau_\text{d}}
\newcommand{\ti}{t_\text{i}}
\newcommand{\tf}{t_\text{f}}

\newcommand{\tpe}{\tau_\text{\tiny P}}
\newcommand{\Ta}{T_\text{\tiny A}}

\newcommand{\g}{\gamma}

\newcommand{\Tf}{T_\text{eff}}
\newcommand{\po}{p_\text{on}}

\newcommand{\tc}{\tilde{\chi}}
\newcommand{\tC}{\tilde{C}}

\newcommand{\Fc}{F_{\textrm{cell}}}

\newcommand{\e}{\varepsilon}

\newcommand{\kp}{k_\text{\tiny P}}
\newcommand{\bp}{b_\text{\tiny P}}
\newcommand{\bii}{b_\text{i}}
\newcommand{\bff}{b_\text{f}}

\newcommand{\ki}{k_\text{i}}
\newcommand{\kf}{k_\text{f}}

\newcommand{\bef}{\beta_\text{f}}
\newcommand{\tp}{\tau_\text{t}}
\newcommand{\Up}{U_\text{\tiny P}}
\newcommand{\Vp}{V_\text{\tiny P}}
\newcommand{\Ap}{a_\text{\tiny P}}
\newcommand{\Fg}{F_\text{\tiny N}}
\newcommand{\Wqsq}{W_\text{\tiny Q}}
\newcommand{\Wqsh}{W_\text{\tiny H}}
\newcommand{\xT}{x_\text{\tiny T}}
\newcommand{\xA}{x_\text{\tiny A}}
\newcommand{\alp}{c_{+}}
\newcommand{\alm}{c_{-}}

\newcommand{\tpp}{\tau_+}
\newcommand{\tmm}{\tau_-}
\newcommand{\Poff}{P_\text{off}}
\newcommand{\Pon}{P_{\text{on}}}
\newcommand{\rh}{p}
\newcommand{\DT}{D_\text{\tiny T}}
\newcommand{\ca}{\chi_\text{\tiny A}}
\newcommand{\caa}{\chi_{\text{\tiny A}1}}
\newcommand{\cab}{\chi_{\text{\tiny A}2}}
\newcommand{\cac}{\chi_{\text{\tiny A}3}}
\newcommand{\cad}{\chi_{\text{\tiny A}4}}
\newcommand{\cia}{\chi_1}
\newcommand{\cib}{\chi_2}
\newcommand{\cic}{\chi_3}
\newcommand{\cid}{\chi_4}

\newcommand{\ta}{t_\text{a}}
\newcommand{\tb}{t_\text{b}}
\newcommand{\tcc}{t_\text{c}}
\newcommand{\tdd}{t_\text{d}}
\newcommand{\EQa}{E_{\text{\tiny Q}1}}
\newcommand{\EQb}{E_{\text{\tiny Q}2}}

\newcommand{\fig}{FIG.}
\newcommand{\eq}{Eq.}

\providecommand{\avg}[1]{\left \langle #1 \right \rangle}
\providecommand{\avgs}[1]{\left \langle #1 \right \rangle_{\text{\tiny SS}}
\providecommand{\avgsh}[1]{\left \langle #1 \right \rangle_{\text{\tiny H}}^}}

\providecommand{\pnt}[1]{\left ( #1 \right)}
\providecommand{\brt}[1]{\left [ #1 \right]}
\providecommand{\abs}[1]{\left | #1 \right|}
\providecommand{\f}[2]{\frac{ #1}{#2}}

\begin{document}

\title{Energetics of active fluctuations in living cells}
\author{\'E. Fodor}
\email[Corresponding author:]{etienne.fodor@univ-paris-diderot.fr}
\affiliation{Laboratoire Mati\`ere et Syst\`emes Complexes, UMR  7057 CNRS/P7, Universit\'e Paris Diderot, 10 rue Alice Domon et L\'eonie Duquet, 75205 Paris cedex 13, France}

\author{K. Kanazawa}
\affiliation{Yukawa Institute for Theoretical Physics, Kyoto University, Kitashirakawa-oiwake cho, Sakyo-ku, Kyoto 606-8502, Japan}

\author{H. Hayakawa}
\affiliation{Yukawa Institute for Theoretical Physics, Kyoto University, Kitashirakawa-oiwake cho, Sakyo-ku, Kyoto 606-8502, Japan}

\author{P. Visco}
\affiliation{Laboratoire Mati\`ere et Syst\`emes Complexes, UMR 7057 CNRS/P7, Universit\'e Paris Diderot, 10 rue Alice Domon et L\'eonie Duquet, 75205 Paris cedex 13, France}

\author{F. van Wijland}
\affiliation{Laboratoire Mati\`ere et Syst\`emes Complexes, UMR 7057 CNRS/P7, Universit\'e Paris Diderot, 10 rue Alice Domon et L\'eonie Duquet, 75205 Paris cedex 13, France}

\date{\today}

\begin{abstract}
The nonequilibrium activity taking place in a living cell can be
monitored with a tracer embedded in the medium. While microrheology
experiments based on optical manipulation of such probes have become
increasingly standard, we put forward a number of experiments with
alternative protocols that, we claim, will provide new insight into
the energetics of active fluctuations. These are based on either
performing thermodynamic--like cycles in control--parameter space, or
on determining response to external perturbations of the confining
trap beyond simple translation. We illustrate our proposals on an
active itinerant Brownian oscillator modeling the dynamics of a probe
embedded in a living medium.
\end{abstract}

\maketitle

\section{Introduction}

A living cell is a nonequilibrium system which needs to constantly
maintain its activity to preserve an organized structure. Major
contributors to this activity are the molecular motors which generate
forces of the order of a piconewton within the cell. This force
generation is an essential process for life as it is the basis of cell
motility, wound healing and cell division. It is fueled by ATP
hydrolysis, thus being a nonequilibrium process commonly named {\em
active} process. The force is applied by the motors on some polar
self--assembled filaments, such as the actin filaments for myosin
motors. The polarity of these filaments added to the force generation
enable the motors to perform a stochastic directed motion. These
phenomena have been experimentally explored {\it in vivo} with living
cells~\cite{fodor,Bursac} and {\it in vitro}, with reconstituted actin
gels in which molecular motors density can be externally
controlled~\cite{toyota,stuhrmann}. 

One of the major experimental technique which has uncovered the
nonequilibrium behavior of living cells and active gels is
microrheology~\cite{Wilhelm,Gallet10,Bursac,toyota,stuhrmann}. Thanks to the progress of high resolution
microscopy it is now possible to track micron sized probes injected
into complex fluids, including living organisms. In addition, by means
of optical or magnetic tweezers, one can apply a controlled force on
these probes, and measure rheological properties such as complex shear
modulus~\cite{weitz99,Yanai04,Desprat03} or creep function~\cite{Gallet06}. By combining these two
measurements, it has been possible to quantify the extent to which the fluctuation--dissipation theorem (FDT) is violated these systems~\cite{Wilhelm, Mizuno}. So far, the central quantity that has been investigated is
a frequency dependent effective temperature~\cite{Cug,Nir,Joanny,Betz,GalletSM}, which serves as an all--purpose measurement of the distance from thermal equilibrium.

Our aim in this paper is to put forward other quantities that can
reveal interesting properties of nonequilibrium activity, and that can
be measured with the same experimental toolbox of microrheology.
In order to render the presentation of these methods more concrete, their predictions shall be illustrated on a recent theoretical model~\cite{fodor}
describing the dynamics of a probe in an active medium.

We begin with giving the basic physical ingredients of our model in section~\ref{sec:model}. We then discuss the simplest protocols in which
the spring constant of a harmonic external potential is changed with
time in section~\ref{sec:spring}. In section~\ref{sec:cycle}, we use a quartic potential
for which two parameters are changed in time to mimick a thermodynamic
cycle~\cite{Kanazawa2}. In section~\ref{sec:response}, we review an already proposed
method of extracting correlations between active force and
position~\cite{Bohec} by exploiting the extended
fluctuation--dissipation relations~\cite{Maes}. In
section~\ref{sec:dissipation}, we apply the Harada--Sasa relation to quantify the dissipation rate arising from the nonequilibrium behavior of the probe~\cite{HarSas}.

\section{Model}
\label{sec:model}
\begin{figure}
\includegraphics[width=\columnwidth]{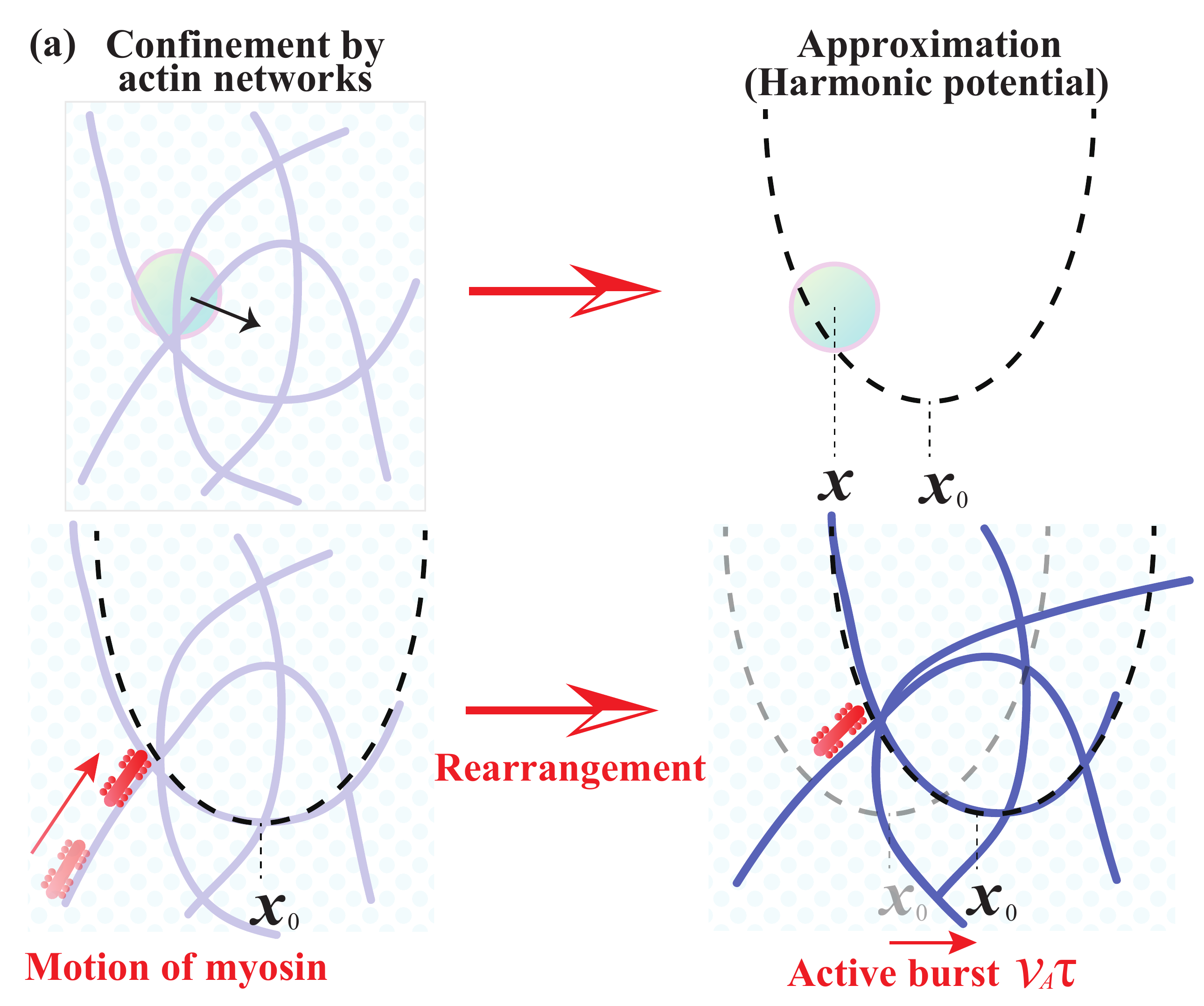}
\includegraphics[width=\columnwidth]{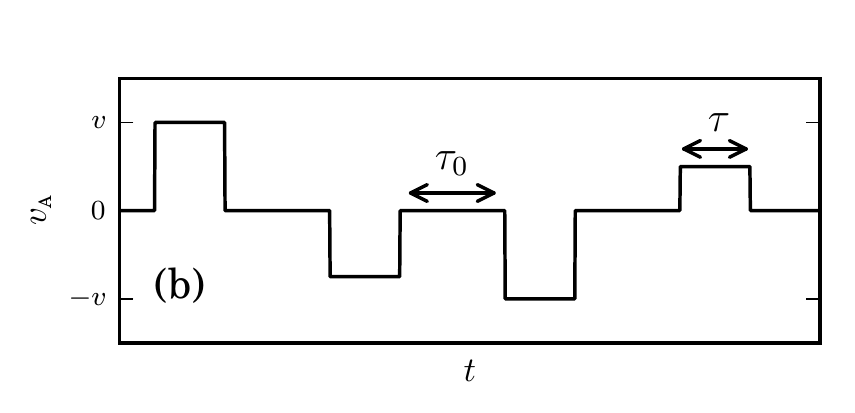}
\caption{\label{fig:act}
(a)~Schematic representation of the energetic landscape
  rearrangement due to motors activity and its modeling using the
  active burst applied on the local minimum. 
In the passive case without motors, the tracer is confined within a harmonic potential. When motors are introduced, their activity modifies the network structure, thus leading to a displacement $\va\tau$ of the potential local minimum $x_0$.
(b)~Example trajectory of the active burst projection $\va$. It equals zero over a random
  duration of average $\tau_0$, and is a random value between $-v$ and $v$ during a random time of order $\tau$.
}
\end{figure}

We model the dynamics of the tracer's position $\br$ by means of an
overdamped Langevin equation as described in~\cite{fodor}. From a
physical viewpoint, the active medium has a complex polymer
cross--linked reticulated structure, surrounded by a viscous Newtonian
fluid. The complex structure of the network confines the particle, and we model
this as a harmonic potential acting on the probe, centered at position
$\br_0$. Active forces which originates from surrounding molecular motors continuously modify
the network structure, thus spatially translating the minimum of this
potential. However, the bead itself modifies the internal network dynamics: arbitrarily large local deformations are unlikely. To account for this feedback mechanism, we introduce a small back action force
on the potential location. Since the harmonic trap models the
confinement by the network, the characteristic size of the trap is much
larger than the particle size to avoid any escape of the particle as shown in \fig~\ref{fig:act}(a). The
back action force is then necessarily small compared to the force
applied on the particle, a feature which we will have to verify in actual experiments. In other words, the tracer dynamics has only
a small effect on the $\br_0$ dynamics, and was in fact neglected
in~\cite{fodor}.
Moreover, the thermal fluctuations applied on the potential center position $\br_0$ must be taken into account, and the corresponding fluctuation amplitude should be negligible compared with the ones of thermal force applied on the tracers.
Introducing a dimensionless parameter $\e\ll1$, which, we anticipate, will be small, we arrive at
the coupled set of equations:
\begin{subequations}
\label{eq:3d}
\begin{align}
 \frac{\dd \brr}{\dd t} & =-\f{1}{\td} (\brr-\brr_0) +\sqrt{\DT}\bxi\,\,, \\
\frac{\dd \brr_0}{\dd t} & =-\frac{\eps}{\td} (\brr_0-\brr) +\bva+\sqrt{\e\DT}\bxi_0\,\,,
\end{align}
\end{subequations}
where $T$ is the bath temperature, $\gamma$ is the friction coefficient of the tracer particle with the surrounding environment, $k$ is the spring constant of the harmonic trap, $\DT=\kb T/\g$ is a thermal diffusion coefficient, and $\td=\g/k$ is a microscopic time scale. The Gaussian white noises $\bxi$ and $\bxi_0$  accounting for thermal
fluctuations are uncorrelated, and $\bva$ is another noise term, referred to as an {\it active burst}, describing the effect
of molecular motors on the network structure. It denotes the velocity at
which the potential is moving, and we model it as a stochastic process
inspired from the dynamics of individual motors: there are quiescent
periods of random duration of average time $\tau_0$ alternating with
active bursts of typical velocity $v$ in a random direction and for
a random time of average $\tau$. In the absence of active forces, this is the itinerant oscillator model introduced by Hill~\cite{0370-1328-82-5-309} and Sears~\cite{0370-1328-86-5-306} within the framework of simple liquids dynamics (see \cite{doi:10.1142/9789814355674} for a review) which has equilibrium dynamics.
Such  dynamics for the tracer particles is associated with a complex modulus of the form~\cite{mason00,Mason}: $G^*=i\omega\eta (1+\eps+i\omega\td)/(\eps+i\omega\td)$. The viscosity $\eta$ is related to the friction coefficient $\g$ via Stokes' law: $\g=6\pi a \eta$, where $a$ is the tracers' radius. Within this minimal rheology, we assume the material behaves like a fluid at short and large time scales, with associated viscosity $\eta$ and $\eta/\e$, respectively, to leading order in $\e$. Thus, this material behaves like a much more viscous fluid at large time scale compared with the short time scale behavior.
In experimental measurements, one has direct access to one dimensional projections of the
position. We shall thus look at the one dimensional projection of
\eq~\eqref{eq:3d} on the scalar position $x$:
\begin{subequations}\label{eq-model}
\begin{eqnarray}\label{eq:modx}
\f{\dd x}{\dd t} &=& -\f{1}{\td}(x-x_0)+\sqrt{\DT}\xi
\,\,,
\\ 
\f{\dd x_0}{\dd t} &=& -\f{\e}{\td}(x_0-x)+\va+\sqrt{\e\DT}\xi_0
\,\,,
\end{eqnarray}
\end{subequations}
where $\avg{\xi(t)\xi(t')}=\delta(t-t')=\avg{\xi_0(t)\xi_0(t')}$ are
still Gaussian noises, and $\va$ equals $0$ over a random duration of
order $\tau_0$ and is a uniform random value between $-v$ and $v$ over
duration of average $\tau$, as depicted in \fig~\ref{fig:act}(b). The
active burst projection $\va$ is a non--Gaussian
process~\cite{Kanazawaprl,Kanazawa}, and the $2$--time correlation
function reads: $\avg{\va (t) \va(0)}=\kb\Ta
\ee^{-|t|/\tau}/(\tau\g)$. The energy scale $\kb \Ta$ defines an
effective active temperature in terms of the duty ratio
$\po=\tau/(\tau+\tau_0)$:
\begin{equation}
\kb \Ta=\f{\g v^2\tau\po}{3}
\,\,.
\end{equation}
It quantifies the amplitude of the active fluctuations, as defined by the active force correlations, and we shall see it characterizes the tracer's statistics at large time scale.
We postpone to appendix~\ref{app:va} the derivation of the $n$--time correlation function of the active burst $\va$. We derive the analytic expressions of the physical observables to leading order in $\e$.

\begin{figure}
\includegraphics[width=0.97 \columnwidth]{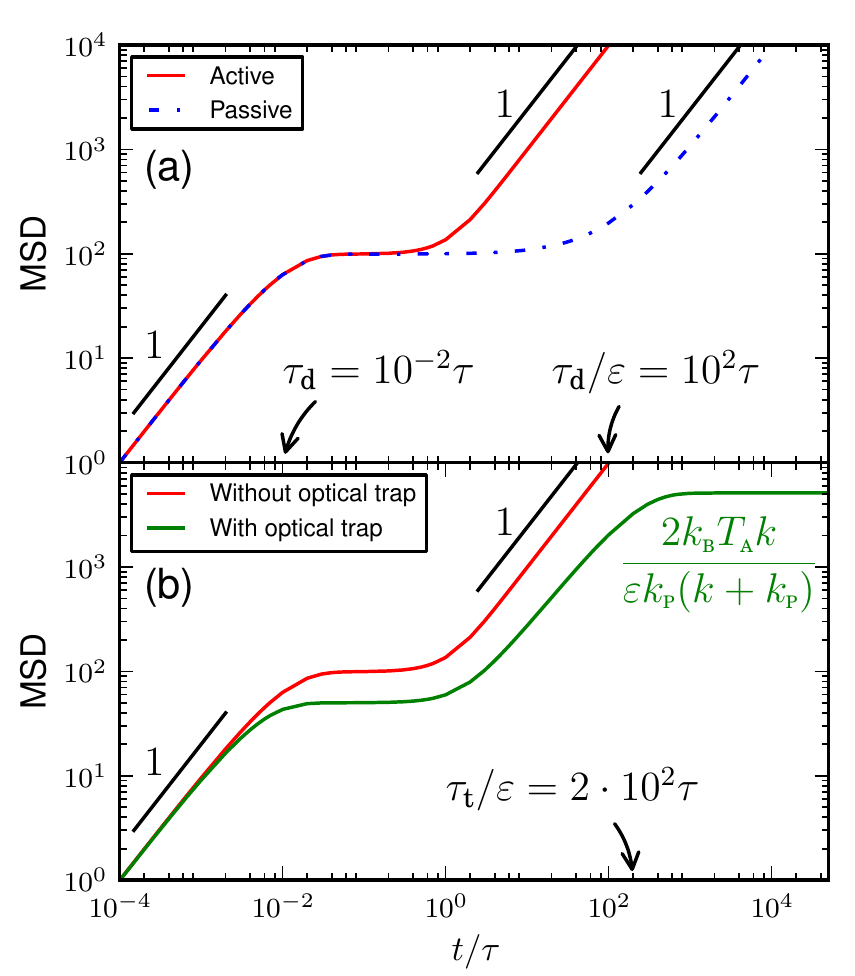}
\caption{\label{fig:msd}(a)~Mean square displacement as a function of the scaled time $t/\tau$
for active (red) and passive (blue) sytems. 
(b)~Mean square displacement as a function of the scaled time $t/\tau$
with (green) and without (red) an external potential, in this case a harmonic
optical trap. The evolution is qualitatively similar for time scale
smaller than $\tp/\e$. At large time scale, there is a plateau due to
the confinement within the optical trap, which value diverges with
$\e$. 
$\{\kb T,\eps,k,\td,\tp,\kb \Ta,\tau\}=\{10^2,10^{-4},2,10^{-2},2\cdot10^{2},1,1\}$.}
\end{figure}

To describe the phenomenology of this model, we focus on the mean
square displacement (MSD) $\avg{\Delta x^2}(\ti,\tf)=\avg{\pnt{x(\ti)-x(\tf)}^2}$. Even though the MSD depends on two time
variables, in the limit where the initial time $\ti$ is large enough
compared to the microscopic relaxation time scale $\td$, it becomes
effectively a function of the only time lag $t=\tf-\ti$.  This is the
case we shall consider in this paper, as we only consider
quasi--static transformations.
Using the Fourier transform of \eq~\eqref{eq-model}, we compute the position autocorrelation function $C(t)=\avg{x(t)x(0)}$, from which we deduce the MSD as: $\avg{\Delta x^2}(t)=2(C(0)-C(t))$. We denote the thermal contribution to the MSD by $\avg{\Delta \xT^2}$, and the MSD when the particle is only subjected to motor activity by $\avg{\Delta \xA^2}$.
We compute these two contributions to leading order in $\e$:
\begin{subequations}
\begin{eqnarray}
\avg{\Delta \xT^2} (t) &=& \f{2\kb T}{k} \pnt{ 1 - \ee^{-t/\td} + \eps \f{t}{\td} }
\,\,,
\\
\avg{\Delta \xA^2} (t) &=& \f{2\kb\Ta/k}{1-(\tau/\td)^2} \bigg[ \ee^{-t/\td} + \f{t}{\td} -1
\nonumber
\\
& & +\pnt{\f{\tau}{\td}}^3\pnt{1 - \ee^{-t/\tau} - \f{t}{\tau}}\bigg]
\,\,.
\end{eqnarray}
\end{subequations}
In the active case, the time evolution
of the MSD exhibits a two step growth with an intermediate plateau. The first growth and saturation
correspond to the equilibrium--like behavior of a probe caged in a
fixed trap. The initial growth is diffusive with a standard diffusion
coefficient $D_\text{\tiny T}$, and the plateau value is given by $2\kb T/k$.
The evolution of the MSD at larger time scales, reflecting the nonequilibrium features of the
system, is a diffusive growth with a diffusion coefficient $\eps D_\text{\tiny T}+D_\text{\tiny A}$, where $D_\text{\tiny A}=\kb\Ta/\g$ is an ``active'' diffusion coefficient.
In the passive case, the tracer particle can also escape the confinement at time scales larger $\td/\eps$, and
the large time scale diffusion coefficient $\e D_\text{\tiny T}$ is small compared to the short time scale one, as shown in \fig~\ref{fig:msd}(a).
The expression of the thermal diffusion coefficient at large time scale agrees with the fluid--like behavior of the material with the associated viscosity $\eta/\e$.
The back action reflects the ability of the particle to modify its environment. The local minimum motion is
not only affected by activity within the network, but also by the
interaction of the bead with the network. The large time scale diffusion
in the passive case is in agreement with experimental observations of
tracers embedded in living cells~\cite{fodor,toyota}. Assuming $\td\simeq1$~ms
and given large time scale diffusion appears for $t>10$~s in~\cite{fodor}, we deduce
$\e\simeq10^{-4}$ in agreement with $\e\ll1$.

\section{Varying the spring constant}
\label{sec:spring}
One of the most fruitful approaches to gather information in living
cells has been achieved by applying external forces to probe
particles. This has been carried out by different methods, such as optical or
magnetic tweezers~\cite{Wilhelm,guo13}, resulting in an effective external
potential $\Up$ acting on the probe. To our knowledge, the general protocol has always been
to apply the potential and then to execute a space translation, typically
with an oscillation, to measure quantities such as the complex shear
modulus. Here we would like to pursue a different route, where, instead of
translating the potential well, we consider a time--dependent change in other parameters of
the external potential. Our main goal is to design a protocol
with time--dependent parameters and to estimate the
work extracted over the whole protocol. Optical tweezers effects are
well approximated by a harmonic potential, though more complex energy landscape can be crafted~\cite{Ciliberto}.

\begin{figure}
\includegraphics[width=\columnwidth]{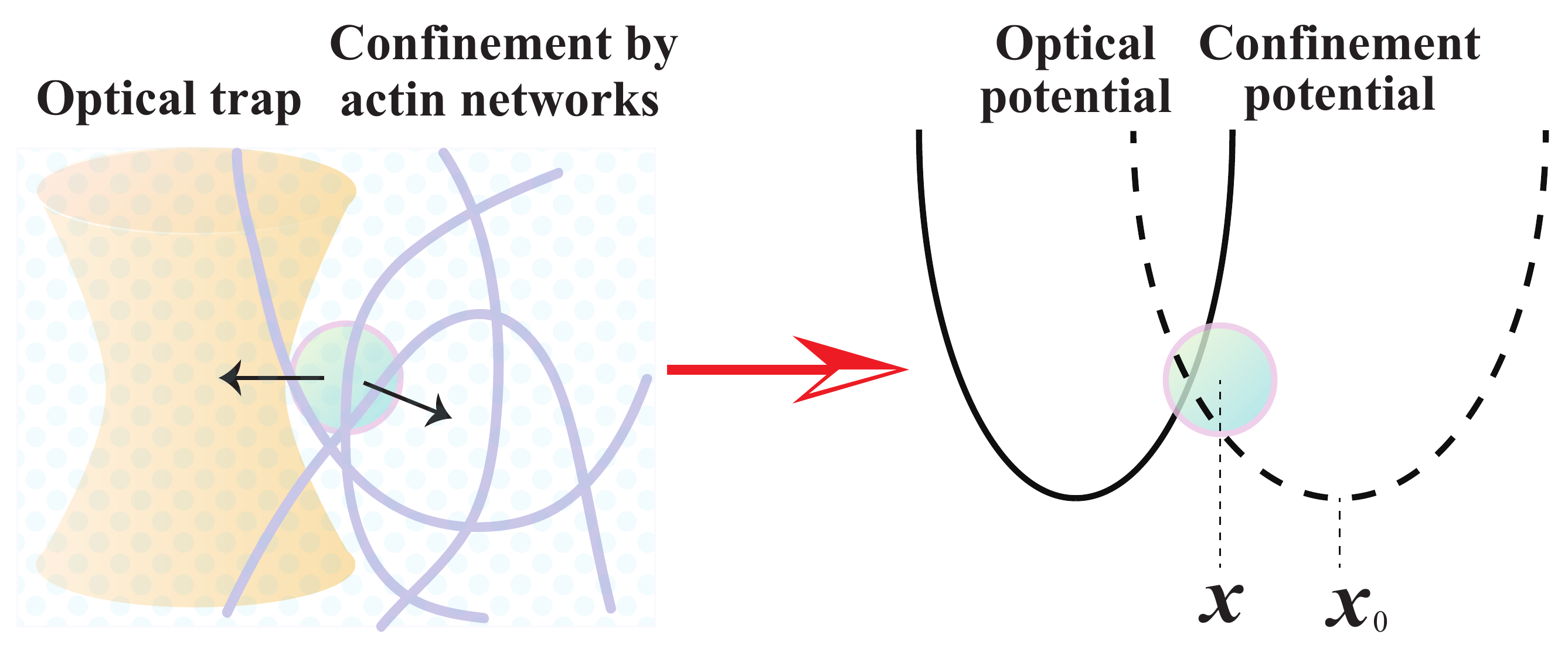}
\caption{\label{fig:sc}
Schematic representation of the energetic landscape when a quadratic optical trap is applied on the tracers, in addition to the harmonic confinement potential.
}
\end{figure}

The simplest protocol is thus to slowly vary the spring constant $\kp$
in time. We consider an external potential $\Up=\kp x^2/2$ is applied
to the tracer as presented in \fig~\ref{fig:sc}, so that an additional
term $-\kp x/\g$ is to be inserted in the $x$ dynamics in
Eq.~\eqref{eq:modx}. We postpone the derivation of the MSD to
appendix~\ref{app:x2}.  Within our model, when we apply this external
force, the evolution of the MSD for time scales smaller than $\tp/\e$,
where $\tp=\g(k+\kp)/(k\kp)$ to leading order in $\e$, is
qualitatively similar to the case without optical trap. At large time
scale, the MSD saturates meaning the tracer is confined within the
optical trap. After a relaxation time $\tp/\e$, the system reaches a
steady state characterized by active fluctuations, the optical trap
stiffness, and the properties of the network via $k$ as presented in
\fig~\ref{fig:msd}(b). Note that the plateau value $2\kb\Ta
k/(\e\kp(k+\kp))$ does not depend on the bath temperature $T$ to
leading order in $\e$, and it diverges with $\e$ so that the back
action is necessary to model the confinement of the bead by the
optical trap.  We show that the stationary displacement probability
density function is a Gaussian distribution to leading order in $\e$,
so that the non--Gaussian nature of the active process $\va$ does not
affect the steady state tracer's distribution to that order of the
calculation. Likewise, the leading term in $\e$ of the tracer's
stationary distribution is unchanged when considering a white noise
for $\va$, be it Gaussian or not.  The time scales $\tau$ and $\tau_0$
do appear to the next orders in $\e$ of the steady state distribution
though.  To quantify the deviation of the stationary distribution from
a Gaussian distribution, we determine the non--Gaussian parameter
(NGP):
\begin{equation}
\text{NGP} = \f{\avgs{x^4}}{3\avgs{x^2}^2} -1 
\,\,,
\end{equation}
where $\avgs{\cdot}$denotes the steady state average. The NGP is zero
for a Gaussian distribution and is often used to quantify deviations
to the Gaussian distribution~\cite{rahman}. We compute this quantity to
leading order in $\e$, as presented in appendix~\ref{app:x2}:
\begin{equation}
\text{NGP} = \f{2\e}{5(1+k/\kp)} \f{9\tau_0^2+3\tau \tau_0-\tau^2}{(\tau+\tau_0)\td}
\,\,.
\end{equation}
The NGP is proportional to $\e$, as another evidence
that the tracer's statistics is Gaussian to leading order in $\e$.
As far as the active temperature $\Ta$ is concerned, it can be determined
independently of the active time scales by applying a quadratic
external potential on the tracer, and by measuring its stationary
distribution of displacement.  It can also be measured from the large
time scale diffusion in the absence of external potential. The method
we propose is more convenient because the tracer does not experience
large excursions, which would otherwise make it hard to keep in focus,
as it remains confined within the optical trap.

\begin{figure}
\includegraphics[width=0.97 \columnwidth]{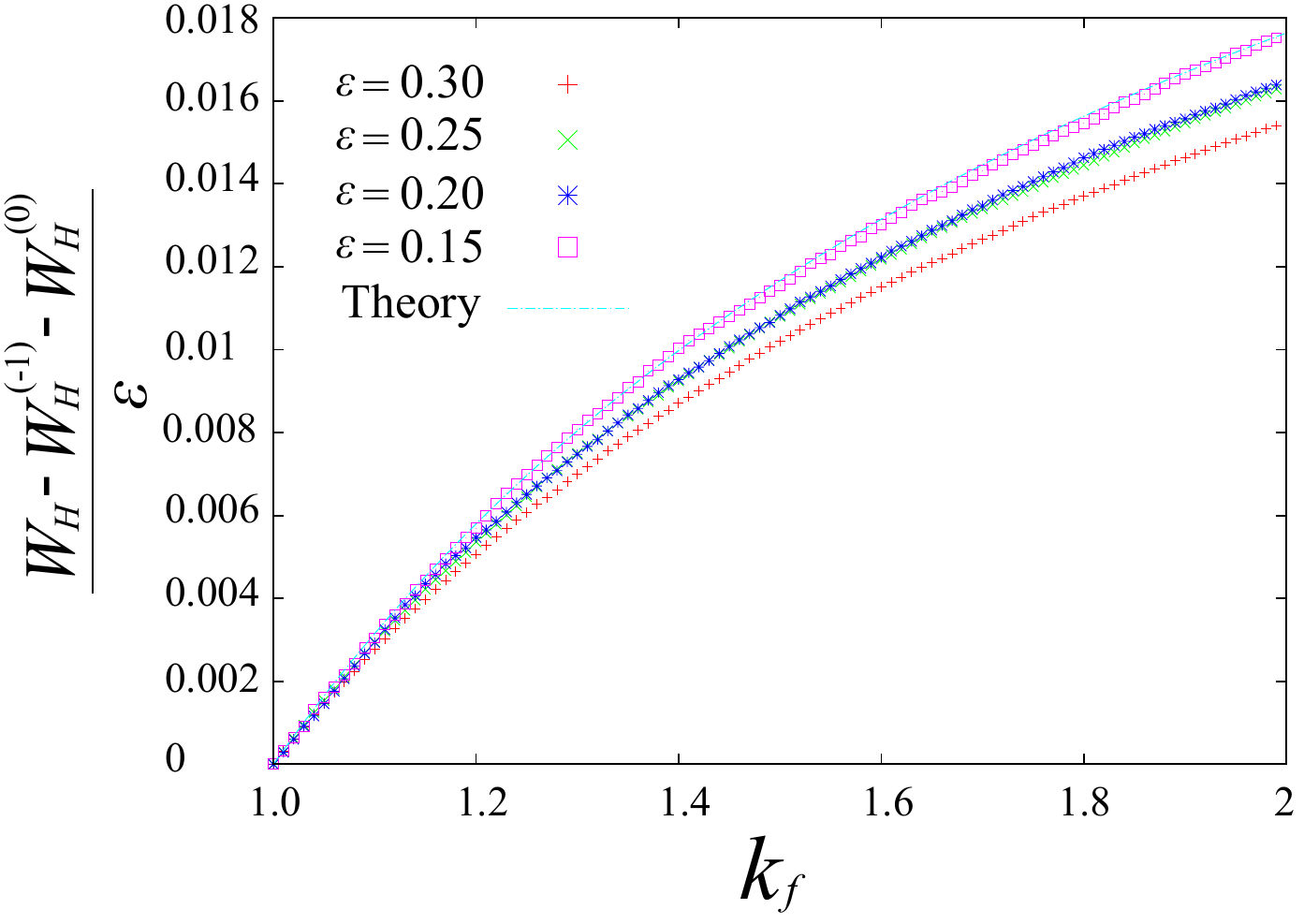}
\caption{\label{fig:sim}
Study of the influence of the $\cO(\e)$ correction term in Eq.~\eqref{work}.
The quasistatic work $W_\text{\tiny H}$ is obtained numerically from simulations of the dynamics in Eq.~\eqref{eq-model}, where $\e=\{0.3,0.25,0.2,0.15\}$. We extract the correction term as: $W_\text{\tiny H}-W_\text{\tiny H}^{(-1)}-W_\text{\tiny H}^{(0)}$, where the expression of $W_\text{\tiny H}^{(n)}=\cO(\e^n)$ is given by Eq.~\eqref{work}. The analytic expression of the $\cO(\e)$ correction term is plotted in cyan dotted line as a function of $\kf$, and it agrees with numerical simulations for $\e=0.15$. For larger values of $\e$, the next order terms should be taken into account to explain for the deviation of the simulated quasistatic work from the prediction in Eq.~\eqref{work}.
$\{T,k,\ki,\g,\tau_0,\tau,v\}=\{0,1,1,1,5,0.6,4\}$}
\end{figure}

The back action reflects the ability of the tracer to act on the
surrounding network, thus affecting the dynamics of the local minimum. In
the present case, it exerts a force on the network which compensates the
driving force due to the active burst, so that a work is applied by
the tracer on the network. We shall see that the measurement of this work
enables one to characterize activity within the system.
We consider a protocol where $\kp$ is slowly varied from $\ki$ to
$\kf$, that is the time evolution of the protocol is much longer than
$\tp/\e$, and the time variation of $\kp$ is negligible compared to
the inverse duration of the protocol in terms of $\e$. The quasistatic
work $\Wqsh$ done by applying the external potential to the probe
is~\cite{Sasa,Sekimoto}:
\begin{equation}\Wqsh=\f{1}{2}\int\dd\kp\avgs{x^2}
\,\,,
\end{equation}
 where the $\avgs{\cdot}$ means that the average is taken in the
steady state with a {\em fixed} optical trap, in the present case a harmonic trap of constant
$\kp$. We determine the expression of this quasistatic work in appendix~\ref{app:x2}.
It takes the form: $W_\text{\tiny H}=E_\text{\tiny H}(\kf)-E_\text{\tiny H}(\ki)$, where $E_\text{\tiny H}$ reads:
\begin{eqnarray}\label{work}
E_\text{\tiny H}(\kp) &=& \f{\kb\Ta}{2\e}\ln\brt{\f{\kp}{k+\kp}} - \f{k\kb\Ta}{2(k+\kp)} + \f{\kb T}{2}\ln\brt{\kp}
\nonumber
\\
& &+ \f{\kb\Ta}{2} \pnt{\f{\tau}{\td}}^2 \ln \brt{ \f{\kp\tau+k(\tau+\td)}{k+\kp} } 
\nonumber
\\
& &- \f{\kb\Ta}{2}\ln\brt{\f{\kp}{k+\kp}}+ \cO(\e)
\,\,.
\end{eqnarray}
This energy scale is defined up to a constant which should render the
argument of the logarithms dimensionless.  It diverges with $\e$,
meaning that if the back action mechanism were neglected it would take
an infinite work to confine the tracer in a harmonic well.  We have
run numerical simulations to determine the accuracy of the above
formula. There is a perfect agreement with our prediction for small
values of $\e$.  When $\e\simeq0.15$, the term of order $\e$ in
Eq.~\eqref{work} is no longer negligible. We compute the expression of
the $\cO(\e)$ correction term, and we show it indeed explains for the
deviation of numerical results with Eq.~\eqref{work}, as presented in
\fig~\ref{fig:sim}. Note that in the passive case, without active
bursts, the work does not vanish but reduces to the difference of the
Helmholtz free energies, as it should for an adiabatic isothermal
transformation. This contribution enters in the $\cO(1)$ term of the
above formula. An interesting feature of formula~\eqref{work} is that
the work is independent of $T$ to leading order in $\e$, meaning that
it should be possible to directly access $\Ta$ with a rather simple
protocol. For example, one could measure the average work with
different values of $\ki$ and $\kf$ to deduce values for $k$, $\eps$
and $\Ta$. However, one should be aware the protocol has to be
operated over large time scales to remain quasistatic.  If the
operator reduces the volume accessible by the bead by setting
$\kf>\ki$, the work is positive, in agreement with the fact that the
probe ``cools down'' when $\kp$ increases. Considering a circular
protocol for which $\kf=\ki$, the extracted work is zero as for an
equilibrium process. The nonequilibium properties remain hidden for a
circular and adiabatic protocol when a harmonic trap is applied to a
tracer.

\section{Thermodynamic cycles with quartic potentials}
\label{sec:cycle}

By combining multiple optical tweezers it is possible to confine the
tracer in a more complex potential such as a double well~\cite{Ciliberto}. The corresponding quartic
optical trap $\Up=\kp x^2/2+\bp x^4/4$ depends on two parameters that are both tunable
by the operator. In particular, the parameter $\kp$ can take negative values, as long as the condition $k+\kp>0$ is fulfilled. We regard the potential anharmonicity as a
small perturbation with respect to the harmonic case: $\bp=\cO\pnt{\e^n}$.
Our picture is that $\e$ is a material--dependent quantity, but the shape of the trap, namely the parameter $n$, is fully controlled by the operator. We consider a quasistatic protocol where $\kp$ varies as before and $\bp$ is set constant. The associated
work is expressed as: $\Wqsq=\int\dd\kp\avgs{x^2}/2$. The steady state average is different from the value presented before due to the quartic term in the optical trap.
By using a perturbation method with respect to $\bp$, we derive the expression of this steady state average to order $\bp$. It follows the quasistatic work from an initial value $\ki$ to a final one $\kf$ is expressed as: $\Wqsq=\Wqsh+\EQa(\kf,\bp)-\EQa(\ki,\bp)+\cO(\bp^2)$, where $\EQa$ is linear in $\bp$. We compute the expression of $\EQa$ to leading order in $\e$, as presented in appendix~\ref{app:x4}:
\begin{eqnarray}\label{eq:EQa}
\f{\EQa(\kp,\bp)}{\bp} &=& \pnt{\f{\kb\Ta}{2k\e}}^2 \bigg[ \f{2 k^4}{\kp^2(k+\kp)^2} +\f{3k^2\tau}{\kp^2(\tau+\td)}
\nonumber
\\
& &- 6k\tau\pnt{ \f{3\tau+2\td}{\kp(\tau+\td)^2} + \f{1}{\td(k+\kp)} }
\nonumber
\\
& &+ \f{6\tau^5}{\td^2(\tau+\td)^3}\ln\brt{k(\tau+\td)+\kp\tau} 
\nonumber
\\
& &-\f{6\tau(6\tau^2+8\tau\td+3\td^2)}{(\tau+\td)^3}\ln\brt{\kp}
\nonumber
\\
& &-\f{6\tau(\tau-3\td)}{\td^2}\ln\brt{k+\kp} \bigg] + \cO(1/\e)
\,\,.
\nonumber
\\
\end{eqnarray}
As for $E_\text{\tiny H}$, it is defined up to constant.
The contribution $\EQa$ of the quartic term in the quasistatic work is of order
$\e^{n-2}$ to leading order in $\e$. Given this contribution should be negligible with respect to $\Wqsh$, we deduce $n$ should fulfil the condition $n>1$.

\begin{figure}
\includegraphics[width=0.97 \columnwidth]{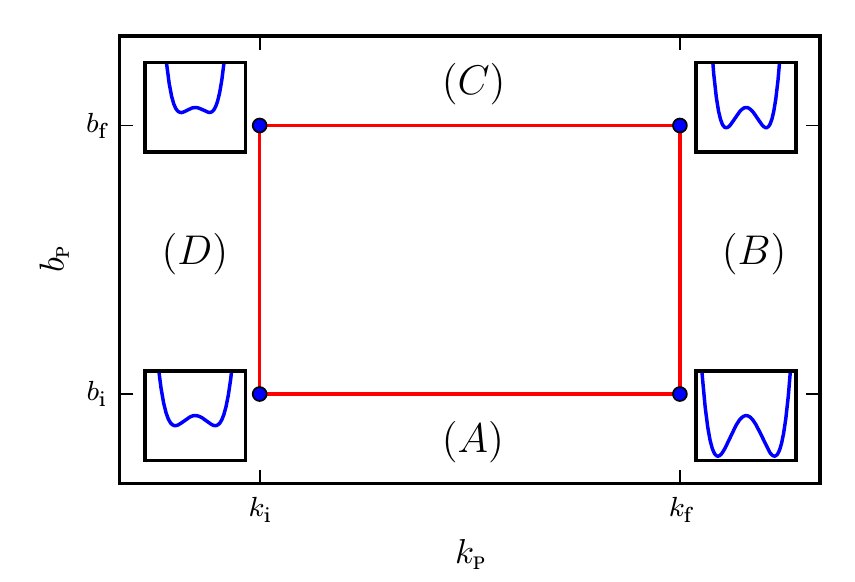}
\caption{\label{fig:circ}Schematic representation of cycle $\C$. The optical trap parameters undergo the transformations $(A)$ to $(D)$: $\{\bii,\ki\}\overset{(A)}{\longrightarrow}\{\bii,\kf\}\overset{(B)}{\longrightarrow}\{\bff,\kf\}\overset{(C)}{\longrightarrow}\{\bef,\ki\}\overset{(D)}{\longrightarrow}\{\bii,\ki\}$. The shape of the external potential tuned by the operator is depicted in blue as a function of the position for the four parameter sets, where $\kf=2\ki<0$ and $\bff=2\bii$.}
\end{figure}

We consider a circular protocol $\C$ where both $\kp$ and $\bp$ are modified in time.  The simplest protocol is then given by four elementary transformations during which
a single parameter is varied, the other one remaining constant.  The
cycle is illustrated in \fig~\ref{fig:circ}. It connects four points in the $\{\kp,\bp\}$ plane:
\begin{eqnarray}
\{\bii,\ki\}\overset{(A)}{\rightarrow}\{\bii,\kf\}\overset{(B)}{\rightarrow}\{\bff,\kf\}\overset{(C)}{\rightarrow}\{\bff,\ki\}\overset{(D)}{\rightarrow}\{\bii,\ki\}
\,\,.
\nonumber
\\
\end{eqnarray}
The associated quasistatic work is defined as:
\begin{equation}
W_{\C}=\f{1}{2}\oint_{\C}\dd\kp\avgs{x^2}+\f{1}{4}\oint_{\C}\dd\bp\avgs{x^4}
\,\,.
\end{equation}
To leading order in $\bp$, the steady state average $\avgs{x^4}$ in the above formula is evaluated 
for a quadratic optical trap, as we compute it in appendix~\ref{app:x2}.
It follows the quasistatic work associated with the protocol $\C$ is expressed to leading order in $\bii$ and $\bff$ as:
\begin{eqnarray}\label{eq:wcirc}
W_\C &=& \EQa(\kf,\bii) - \EQa(\ki,\bii) + \EQb(\kf,\bff) - \EQb(\kf,\bii)
\nonumber
\\
& &+ \EQa(\ki,\bff) - \EQa(\kf,\bff)
\nonumber
\\
& &+ \EQb(\ki,\bii) - \EQb(\ki,\bff)
\,\,,
\end{eqnarray}
where $\EQb$ is linear in $\bp$:
\begin{equation}\label{eq:EQb}
\f{\EQb(\kp,\bp)}{\bp} = 3\pnt{\f{k\kb\Ta}{2\kp(k+\kp)\e}}^2 + \cO(1/\e)
\,\,.
\end{equation}
The formula~\eqref{eq:wcirc} reveals one can measure some work for a
circular protocol if the external potential applied on the tracer
contains an anharmonic component~\cite{Kanazawa2}. The equilibrium
counterpart of this work vanishes, namely for the itinerant
oscillator case when $\Ta=0$, and a nonzero work can thus be regarded as a signature of
nonequilibrium activity within the system. The work applied during
such a protocol is of order $\e^{n-2}$ to leading order in $\e$. Being
$n$ necessarily greater than $1$, we deduce this work is negligible
compared with the work associated to the protocol presented in
section~\ref{sec:spring}. Thus, the anharmonicity of the external
potential leads to a nonzero quasistatic work for a circular protocol,
but its small value may be hard to measure experimentally.  Assuming
experimental apparatus enable one to detect such a work, the active
temperature can then be extracted from this measurement, given the
back action strength $\e$ has been estimated by another method and the
parameter $n$ is controlled by the operator.  A simple method to fix
$n$ is to tune the anharmonicity so that it gives a nonzero
contribution to $\Wqsq-\Ta\ln\brt{\kf(k+\ki)/\ki/(k+\kf)}/(2\e)$, by
detecting when the value of this work differs from the order $\e^0$ in
$W_\text{\tiny H}$. In such a case, the contribution of $\EQa$ is to
be taken into account, so that it corresponds to the case $n=2$.
Note that neither $\EQa$ nor $\EQb$ depend on $\tau_0$ to leading order in $\e$. The waiting time scale $\tau_0$ affects the next order in $\e$ of the work associated with the cycle $\C$. Moreover, the work applied during such a protocol does not vanish in the limit where the active process $\va$ becomes a white noise, namely when $\{\tau,v\}\to\{0,\infty\}$ with fixed $\Ta$. In such a limit and assuming $T=0$, the dynamics presented in Eq.~\eqref{eq-model} describes the evolution of a particle subjected to a white non--Gaussian noise, so that one can indeed extract work from a cycle as already noticed in~\cite{Kanazawa2}.

\section{Effective temperature and force--position correlations}
\label{sec:response}

\begin{figure}
\includegraphics[width=0.97 \columnwidth]{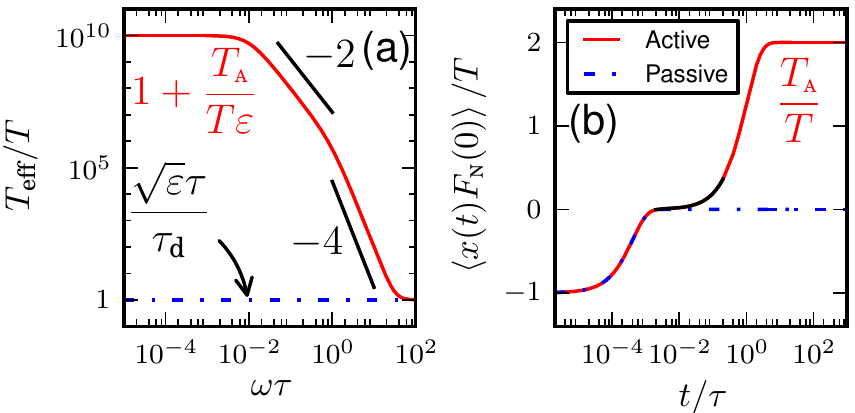}
\caption{\label{fig:xf} 
(a)~Effective temperature as a function of the scaled frequency $\omega \tau$. The plateau value at low frequency equals $T+\Ta/\eps$, and it equals $T$ at high frequency as for the passive case. Between the two saturations, it scales successively like $1/\omega^2$ and $1/\omega^4$, provided the time scales $\tau$ and $\td/\sqrt{\eps}$ are well separated.
(b)~Evolution of the force--position correlation function with the scaled
time $t/\tau$ in the passive (blue) and active (red) cases. The correlation function is negative at short time scale with an initial value $-\kb T$. It remains negative in the passive case. There is a linear growth regime in the active case as depicted in black, and the correlation function saturates to a plateau value $\kb\Ta$.
(a)~$\{T,\eps,\td,\Ta,\tau\}=\{1,10^{-8},1,10^2,10^2\}$.
(b)~$\{T,\td,\Ta,\tau\}=\{5,0.2,10,5\cdot10^2\}$.}
\end{figure}

Active microrheology experiments on living cells measure the response $\chi$ to an
external stress, and its temporal Fourier transform $\tc(\omega)=\int \dd t \ee^{-i \omega t} \chi(t)$. The
latter is, up to a constant, the inverse of the complex modulus
$G^{*}$~\cite{schmidt97}. Following Lau {\it et al.}~\cite{Lau},
the tracer's evolution in a viscous fluid is modeled as:
\begin{equation}
\gamma \frac{\dd x}{\dd t}= \Fc(t) \,\,,
\end{equation}
where $\Fc$ describes all the forces arising from the medium. Within
this minimal assumption several works have
measured the nonequilibrium properties of the force $\Fc$~\cite{Mizuno,gallet09}. These were
quantified by looking at the deviation from such equilibrium relations
as the fluctuation--dissipation theorem. For example, the correlation--to--response ratio leads to a frequency--dependent ``effective temperature'' as~\cite{Cug,Nir,Joanny}:
${\Tf(\omega)=-\omega\tC(\omega)/(2\kb\tc''(\omega))}$, where $\tc''$
is the imaginary part of the response Fourier transform, and $\tC$ is
the position autocorrelation function in the Fourier domain. Of course, this effective temperature is not a {\it bona fide} temperature, in the sense that even in a stationary regime it is generally observable--dependent, but the fact that its high frequency value collapses to the bath temperature in the absence of nonequilibrium processes constitutes a useful benchmark. This is the simplest manner to evaluate the distance from equilibrium.
In the absence of external potential as described in Eq.~\eqref{eq-model}, we compute analytically this temperature to leading order in $\e$: 
\begin{equation}\label{eq:Teff}
\Tf(\omega)=T+\f{1}{\eps+\pnt{\omega \td}^2}\f{\Ta}{1+\pnt{\omega \tau}^2}
\,\,.
\end{equation}
At high frequencies, the effective temperature coincides with the bath
temperature $T$, meaning thermal fluctuations are predominant with respect to motor activity in this regime, in agreement with the MSD short time behavior. The plateau value $T+\Ta/\eps$ at low frequency represents an alternative measurement of the active fluctuation amplitude.
Between the two plateaus, the effective temperature successively scales like $1/\omega^4$ and $1/\omega^2$ given the two time scales $\tau$ and $\td/\sqrt{\eps}$ are well separated as shown in \fig~\ref{fig:xf}(a), thus providing a way to determine these time scale values from the slope variation. 
When we neglect the back action effect, the effective temperature diverges at low frequencies. It results from the fact that the active MSD diffuses at large time scale, whereas it saturates to the equilibrium value for a passive system. The introduction of the back action changes the rheology of the material, so that the passive MSD also diffuses at large time scale, from which we deduce the effective temperature saturates at low frequency.

A generalization of usual microrheology measurements relies on
applying an arbitrary perturbation on the tracers and measuring their
response function. The external stimulus is generally a homogeneous
force. We address here the case where an arbitrary potential $\Vp=-\Ap(t)
V(x(t))$ is applied on the tracers. The generalized tracers' response $\chi_\text{\tiny G}$
quantifies the effect of the perturbation on an arbitrary observable $A$:
\begin{equation}
\chi_\text{\tiny G} (s,u) = \left.\f{\delta \avg{A(s)}}{ \delta \Ap(u) }\right|_{\Ap=0}
\,\,.
\end{equation}
Causality ensures the response function is zero when the
measurement is performed before the perturbation, at
$u<s$. Since the thermal noise has a Gaussian statistics, the probability weight $\cP$ associated with a given realization of the thermal noise is defined as: $\cP\brt{\xi}\propto \ee^{-\cA\brt{\xi}}$, where $\cA\brt{\xi}=\int \dd t' \xi^2(t')/(4\g\kb T)$ is the Onsager--Machlup (or action) functional, in which $\xi$ determines the dynamics of the probe~\cite{Maes,Bohec}. The application of the external potential $\Vp$ results in a variation $\delta\cA$ of the action functional, so that the response function is expressed as:
\begin{equation}
\chi_\text{\tiny G} (s,u) = -\avg{A(s)\left.\f{\delta\cA}{ \delta \Ap(u)} \right|_{\Ap=0} }
\,\,.
\end{equation}
To determine the response function, we only need to compute the action functional to leading order in $\Ap$:
\begin{eqnarray}
\cA = -\int \dd t' \f{ \Ap(t')}{2 \g \kb T}\brt{\g \f{\dd x}{\dd t'} - \Fg(t') } \f{\dd V(x(t'))}{\dd x} + \cO(\Ap^2)
\,\,,
\nonumber
\\
\end{eqnarray}
where $\Fg=-k(x-x_0)$ is the force reflecting the interaction of the tracer with the surrounding actin network. We deduce the response function in terms of the probe's statistics and the network force:
\begin{eqnarray}\label{eq:xf}
\chi_\text{\tiny G} (s,u) &=& \f{1}{2 \g \kb T}\bigg[\g \f{\p C_\text{\tiny AV}(s,u)}{\p u}
\nonumber
\\
& &-\avg{A(s)\f{\dd V(x(u))}{\dd x}\Fg(u)} \bigg]
\,\,,
\end{eqnarray}
where $C_\text{\tiny AV}(s,u)=\avg{A(s)V(u)}$. This expression reveals that one can gain
information about the correlation between the network force and the
tracers' statistics by independently measuring $\chi_\text{\tiny G}$ and $C_\text{\tiny AV}$.

In the case where $\Ap$ is a homogeneous force, when $\Vp=-\Ap x$, the
response function is measured by usual microrheology methods.  If we choose the observable $A$ to be the tracers' position $x$, it is possible to access the force--position correlation
$\avg{x(s)\Fg(u)}$~\cite{Bohec}. After an exponentially fast
initial transient regime which we neglect, this correlation function
depends only on the lag time $t=s-u$. The expression for this correlation is not
invariant under time reversal, and we compute it for the
case $t>0$ to leading order in $\e$:
\begin{eqnarray}\label{eq:xf}
\avg{x(t)\Fg(0)} &=& -\kb T \ee^{-t/\td} + \f{\kb\Ta}{1-(\tau/\td)^2} \bigg[  1-\ee^{-t/\td} 
\nonumber
\\
& & - \pnt{\f{\tau}{\td}}^2 \pnt{1-\ee^{-t/\tau} } \bigg]
\,\,.
\end{eqnarray}
The initial value $-\kb T$ is negative, and equals the thermal fluctuation amplitude in agreement with~\cite{Bohec}. This anticorrelation between the network force and the tracers' displacement is another evidence of the short time scale confinement.  In the active case, the correlation function can take positive values, showing the active burst allows the tracer to overcome the short time scale confinement. When $\tau\gg\td$, there is a linear growth with
coefficient $\kb\Ta/\tau/(1-(\td/\tau)^2)$, and then it
reaches a plateau value $\kb \Ta$ as presented in \fig~\ref{fig:xf}(b).
The linear regime is observed in~\cite{Bohec}, but
the plateau is not present. We speculate that a larger time window would allow
one to observe the saturation of the correlation function. The
existence of the plateau calls for new experiments as it would provide yet another way of measuring the amplitude of active fluctuations. Note that this amplitude is also accessible via the
linear growth coefficient if $\tau$ and $\td$ are already
known. Moreover, a positive value of the force--position correlation function is a signature of nonequilibrium activity within the system as it would remain negative for an equilibrium process.

\section{Energy dissipation and Harada--Sasa relations}
\label{sec:dissipation}

The dissipation within the system is the work applied by the tracer
on the surrounding environment regarded as a heat bath~\cite{Sekimoto}. It has already
been measured in colloidal systems~\cite{ExpJap,ExpAll}, and should be
a good criterion to characterize nonequilibrium activity in biological
systems. We adopt a natural definition for the mean rate of energy
dissipation~\cite{Sasa,Sekimoto}: $J=\avg{v(\g v-\xi)}$, where $v$ is
the velocity of the tracer. It is the difference between the mean
power given by the particle to the heat bath via the drag force $\g
v$, and the one provided in average by the thermostat to the particle
via the thermal force $\xi$. It has been demonstrated by Harada and Sasa that  this quantity is
related to the correlation and response functions defined
previously~\cite{HarSas}: $J=\g\int\dd\omega\brt{\omega\tC(\omega)+2\kb T \tc''(\omega)}\omega/(2\pi)$. This relation presents the heat current $J$ as a quantification of the deviation from the FDT valid for an equilibrium process. Within our model, the energy dissipation rate equals the average power of the network force: $J=\avg{v\Fg}$. We compute it in terms of the microscopic ingredients:
\begin{equation}
J=\f{\kb\Ta}{\tau+\td}
\,\,.
\end{equation}
It is not affected by the back action to leading order in $\e$.
The energy dissipation rate is zero when no activity occurs in the medium for an arbitrary value of $\e$, as expected for an equilibrium process. The dissipation rate depends on the coupling between the probe and its
environment via $\td$.
To minimize the dissipation rate, the
time scale of the quiescent periods $\tau_0$ should be as large as
possible, whereas the time scale of the ballistic jumps $\tau$ should
be very small, in agreement with observations in biological systems
for which $\tau_0>\tau$~\cite{toyota,Mizuno}. As in the previous
section, the definition and the expression of $J$ show one can access
the microscopic features of motor activity via independent
measurements of the correlation and response functions.

\begin{figure}
\includegraphics[width=0.97 \columnwidth]{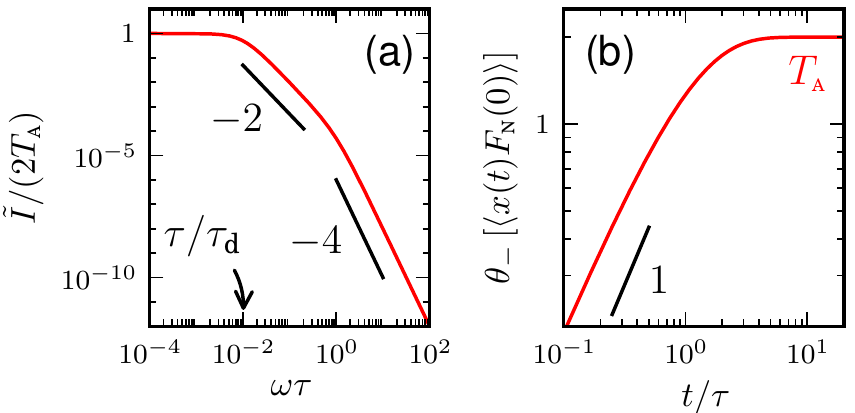}
\caption{\label{fig:I}(a)~Evolution of the Fourier transform of the spectral density of the energy dissipation rate with the scaled frequency $\omega\tau$ when $\tau\gg\td$. The plateau value at low frequency equals $2\kb\Ta$. It scales like $1/\omega^{4}$ at high frequency, and there is a crossover regime $1/\td\ll\omega\ll1/\tau$ with another power law $1/\omega^2$.
(b)~Antisymmetric force--position correlation function as a function of the scaled time $t/\tau$. It is linear in time at short time scale  with a growth coefficient $J$, and saturates to the value $\kb\Ta$ at large time scale. $\{\kb\Ta,\tau,\td\}=\{2,10,0.1\}$.}
\end{figure}

The main drawback of this approach is that one should measure $\tC$
and $\tc$ over a large range of frequencies to access the energy
dissipation rate. Thus, it is interesting to  focus on the
spectral density of the energy dissipation rate:
$\tilde{I}(\omega)=\g\omega\brt{\omega\tC(\omega)+2\kb
T \tc''(\omega)}$, which, when integrated over the whole frequency range,
equals the energy dissipation rate~\cite{HarSas,ExpJap}:
$J=\int\dd\omega \tilde{I}(\omega)/(2\pi)$. To give a physical
interpretation of this quantity, we introduce the operators
$\theta_\pm$ the effect of which on an arbitrary function $f(t)$ is to
extract its even/odd component:
$\theta_\pm\brt{f(t)}=\pnt{f(t)\pm f(-t)}/2$. The Fourier transform of
the symmetrized force--velocity correlation function is
$\tilde{I}$~\cite{HarSas, Baiesi}, so that:
$I(t)=\theta_+\brt{\avg{v(t)\Fg(0)}}$. This relation is a
reformulation of \eq~\eqref{eq:xf} when $V=x=A$, and we shall see
in which sense it enables one to easily access the characteristics of
motor activity. Note that the antisymmetrized force--position
correlation function defined previously is also related to this
quantity: $I(t)=\dd \theta_-\brt{\avg{x(t)\Fg(0)}}/\dd t$. We compute
the dissipation rate spectrum analytically to leading order in $\e$:
\begin{equation}
\tilde{I}(\omega)= \f{1}{1+(\omega\td)^2} \f{2\kb\Ta}{1+(\omega\tau)^2}
\,\,.
\end{equation}
The low frequency plateau provides a direct measurement of the active fluctuation amplitude $\kb\Ta$.  At high frequency, it scales like $1/\omega^{4}$, and there is a crossover regime $\omega_1\ll\omega\ll\omega_2$, where $\omega_1,\omega_2\in\{1/\tau,1/\td\}$, with a power law behavior $1/\omega^{2}$. Thus, one can determine $\tau$ and $\td$ from the variation of the slope, provided the two time scales are well separated, as presented in \fig~\ref{fig:I}(a).
We derive the antisymmetric force--position
correlation function from $\tilde{I}$ to leading order in $\e$:
\begin{eqnarray}
\theta_-\brt{\avg{x(t)\Fg(0)}} &=& \f{\kb\Ta}{1-(\tau/\td)^2} \bigg[  1-\ee^{-t/\td} 
\nonumber
\\
& & - \pnt{\f{\tau}{\td}}^2 \pnt{1-\ee^{-t/\tau} } \bigg]
\,\,.
\end{eqnarray}
It equals the force--position correlation function in Eq.~\eqref{eq:xf} when $T=0$.
At short time scale, growth is linear with a coefficient $J$. Hence, it is possible to estimate the
energy dissipation rate by measuring its spectral density only in the
high frequency domain, which mostly facilitates the experimental task
with respect to the procedure proposed in \cite{HarSas}. The
correlation function saturates to a plateau value $\kb\Ta$ at large
time scale as presented in \fig~\ref{fig:I}(b), showing it provides
an alternative to directly measuring both the energy dissipation
rate and the amplitude of the tracer's active fluctuations.

\section{Conclusion}
\label{sec:conclusion}
We offer theoretical predictions for energetic observables of a system
where both thermal fluctuations and nonequilibrium activity
coexist. We also propose a set of concrete experimental methods and protocols,
so that our predictions may be tested with existing experimental
techniques. These new methods  end up in more stringent constraints on the theoretical modeling which is employed in the studies of tracer dynamics, and thus they should also be a crucial test for the robustness of our own model. By applying such methods, we find that one can access the
microscopic features of motor activity, and fully characterize the
nonequilibrium process arising in the medium. The most natural step
forward is to address the analytic computation of the finite time
extracted work, for which one should find an optimal protocol
maximizing the extracted power~\cite{Crooks}. Another interesting
issue is the excess heat and house--keeping heat produced by such a
protocol, the computation of which requires to determine the
steady--state distribution of the process~\cite{HatSas}. Finally, the
bath temperature could be regarded as another tunable parameter
provided its variation does not modify the microscopic features of the
system~\cite{Sivak}, which is not the case in biological systems but
could be conceivable in colloidal systems.

\acknowledgments A part of the numerical calculations was carried out on SR16000 at YITP in Kyoto University. This work was supported by the
JSPS Core-to-Core Program “Non-equilibrium dynamics of soft matter and information,” the Grants-in-Aid for
Japan Society for Promotion of Science (JSPS) Fellows (Grant No. 24$\cdot$3751), and JSPS KAKENHI Grant No. 22340114.

\appendix
\section{Active bursts' statistics}\label{app:va}

We denote $\Poff$ the transition probability to the state in which $\va$ is zero, and $\Pon$ the transition probability to the state $\va=p v$, where $p$ is a uniform random value between $-1$ and $1$. The set of master equations describing the evolution of the active burst 1D--projection is:
\begin{subequations}
\begin{eqnarray}
\dd_t \Poff(t) &=& \f{1}{\tau}-\Poff(t)\pnt{\f{1}{\tau}+\f{1}{\tau_0}}
\,\,,
\\
\p_t \Pon(t,\rh) &=& \f{\Poff(t)}{2\tau_0}-\f{\Pon(t,p)}{\tau}
\,\,.
\end{eqnarray}
\end{subequations}
We derive the expression of the transition probability $\Pon$ from these equations.  For symmetry reasons, only the $2n$--time correlation functions of the active burst are non--zero. Given the active burst is in the steady state at the initial time, the $2n$--time correlation function $K_\text{\tiny A}\pnt{\{t_i\}}=\avg{\va(t_{2n})\va(t_{2n-1})\dots \va(t_1)}$ reads:
\begin{eqnarray}
\f{K_\text{\tiny A}(\{t_i\})}{v^{2n}} &=& \int\dd^{2n}\rh\Pon^\text{ss}(\rh_1)\rh_1
\prod_{i=2}^{2n}\Pon(t_{i}-t_{i-1},\rh_i|\rh_{i-1})\rh_i
\,\,,
\nonumber
\\
\end{eqnarray}
where $\Pon(t,\rh_b|\rh_a)$ is the transition probability from $\rh_a$ to $\rh_b$, and $\Pon^\text{ss}$ is the steady state transition probability. We deduce the explicit expression of $K_\text{\tiny A}$:
\begin{eqnarray}\label{eq:KA}
K_\text{\tiny A}\pnt{\{t_i\}}=\phi(t_2-t_1)\prod_{i=1}^{n-2} \phi(t_{2i+2}-t_{2i+1})\psi(t_{2i+1}-t_{2i})
\,\,,
\nonumber
\\
\end{eqnarray}
where $t_{2 n}\geq t_{2 n-1}\geq\dots\geq t_1$. The functions $\phi$ and $\psi$ are defined as: 
\begin{subequations}\label{eq:phipsi}
\begin{eqnarray}
\phi(t) &=& \f{v^2 p_\text{on}}{3} \ee^{-\abs{t}/\tau}
\,\,,
\\
\psi(t) &=& 1+ \f{4}{5}\pnt{1+\f{\tau_0}{\tau}} \ee^{-\abs{t}/\tau} + \f{\tau_0}{\tau}\ee^{-\abs{t}\pnt{1/\tau+1/\tau_0}}
\,\,.
\nonumber
\\
\end{eqnarray}
\end{subequations}

\section{Tracers' statistics}
\subsection{Quadratic optical trap}\label{app:x2}
The dynamics of $x$ and $x_0$ is given by the following set of equations:
\begin{subequations}\label{eq:x2}
\begin{eqnarray}
\f{\dd x}{\dd t} &=& -\f{1}{\td}(x-x_0) - \f{1}{\tpe} x+\sqrt{\DT}\xi
\,\,,
\\ 
\f{\dd x_0}{\dd t} &=& -\f{\e}{\td}(x_0-x)+\va+\sqrt{\e\DT}\xi_0
\,\,,
\end{eqnarray}
\end{subequations}
where $\tpe=\g/\kp$. Using the Fourier transform of \eq~\eqref{eq:x2}, we express the tracer's position in terms of the stochastic noises in the Fourier domain as:
\begin{equation}
\tilde{x} = \tilde{\chi} \sqrt{\DT} \tilde{\xi} + \tilde{\chi}_\text{\tiny A} \pnt{ \tilde{v}_\text{\tiny A} + \sqrt{\e\DT} \tilde{\xi}_0 }
\,\,,
\end{equation}
where the functions $\tilde{\chi}$ and $\tilde{\chi}_\text{\tiny A}$ are defined as:
\begin{subequations}\label{eq:chi}
\begin{eqnarray}
\tilde{\chi}(\omega) &=& \f{(\e+i\omega\td)/k}{i\omega\td\pnt{1+\e+i\omega\td} + \kp\pnt{\e+i\omega\td}/k}
\,\,,
\nonumber
\\
\\
\tilde{\chi}_\text{\tiny A} (\omega) &=& \f{1/k}{i\omega\td\pnt{1+\e+i\omega\td} + \kp\pnt{\e+i\omega\td}/k}
\,\,.
\nonumber
\\
\end{eqnarray}
\end{subequations}
Alternatively, the tracer's position is expressed in the time domain as:
\begin{eqnarray}\label{eq:x0}
x(t) &=& \g\int\limits^t \dd t' \bigg[ \chi(t-t') \sqrt{\DT} \xi(t')
\nonumber
\\
& &+ \ca(t-t') \pnt{\va (t') + \sqrt{\e\DT} \xi_0(t')} \bigg]
\,\,.
\end{eqnarray}
By using the residue theorem, we compute from Eq.~\eqref{eq:chi} the expression of $\chi$ and $\chi_\text{\tiny A}$ in the time domain:
\begin{subequations}
\begin{eqnarray}
\chi(t) &=& \f{1}{\g(c_+ - c_-)}\pnt{  c_+ \ee^{-t/\tpp} -  c_- \ee^{-t/\tmm} }
\,\,,
\\
\chi_\text{\tiny A}(t) &=& \f{1}{\g(c_+ - c_-)}\pnt{ \ee^{-t/\tpp} - \ee^{-t/\tmm} }
\,\,,
\end{eqnarray}
\end{subequations}
where $\tau_\pm=\td/(\eps-c_\pm)$, and the coefficients $c_\pm$ read:
\begin{equation}
c_\pm=\f{\eps-1-\kp/k}{2} \brt{1 \pm \sqrt{ 1+ \f{4\e}{\pnt{\eps-1-\kp/k}^2} } }
\,\,.
\end{equation}
We determine the position autocorrelation function in the Fourier domain for an arbitrary $\e$:
\begin{eqnarray}
\tC(\omega) &=& \f{2 (\tpp\tmm)^2/(k\td^3)}{\pnt{1+(\tpp\omega)^2}\pnt{1+(\tmm\omega)^2}} 
\nonumber
\\
& &\times\brt{\pnt{\e + \e^2+(\omega\td)^2} \kb T + \f{\kb\Ta}{1+(\omega\tau)^2}}
\,\,.
\end{eqnarray}
We then deduce the expression of the MSD, without any assumption made on $\e$:
\begin{subequations}
\begin{eqnarray}
\avg{\Delta \xT^2} (t) &=&  \f{2\kb T/k}{(\alp-\alm)(\alm+\alp-2\eps)}
\nonumber
\\
& &\times \bigg[ \f{\alm^2-\eps(1+2\alm)}{\alm-\eps}\pnt{1 - \ee^{-t/\tmm}}
\nonumber
\\
& &- \f{\alp^2-\eps(1+2\alp)}{\alp-\eps}\pnt{1 - \ee^{-t/\tpp}} \bigg]
\,,
\\
\avg{\Delta \xA^2}(t) &=& \f{2\kb\Ta/k}{\pnt{(\tau/\tmm)^2-1} \pnt{(\tau/\tpp)^2-1 } (\alp-\alm)} 
\nonumber
\\
& & \times \bigg[(\alp-\alm) \pnt{1 - \ee^{-t/\tau}}  \pnt{\f{\tau}{\td}}^3 
\nonumber
\\
& & +\f{(\tau/\tpp)^2-1}{\alp+\alm-2\eps} \pnt{1-\ee^{-t/\tmm}}
\nonumber
\\
& & - \f{(\tau/\tmm)^2-1}{\alm+\alp-2\eps} \pnt{1-\ee^{-t/\tpp}} \bigg]
\,\,.
\end{eqnarray}
\end{subequations}
From the saturation value of the MSD at large time scale, we deduce the expression of the steady state average:
\begin{eqnarray}
\avgs{x^2} &=& \f{\kb\Ta k}{\e\kp(k+\kp)} + \f{\kb T}{\kp} 
\nonumber
\\
& &- \f{\kb\Ta k}{(k+\kp)^2} \brt{ \f{\tau}{\td} + \f{k^2(\tau+\td)}{\kp(k+\kp)\tau+k\kp\td} } 
\nonumber
\\
& &+ \cO(\e)
\,\,.
\end{eqnarray}
The expression of $E_\text{\tiny H}$ is given by the primitive of the above formula with respect to $\kp$, thus being defined up to a constant.
To determine the non--Gaussian parameter, we compute the steady state average $\avgs{x^4}$.
Given the tracer's statistics is Gaussian to leading order in $\e$, we can easily deduce $\avgs{x^4}$ to first order in $\e$ from the above formula:
\begin{equation}
\avgs{x^4} = 3\pnt{\f{k\kb\Ta}{\e\kp(k+\kp)}}^2 + \cO(1/\e)
\,\,.
\end{equation}
The computation of the next order requires to develop the expression of $x^4$ in terms of $\chi$ and $\ca$.
From Eq.~\eqref{eq:x0}, we split the steady state average in two contributions:
\begin{equation}
\avgs{x^4} = \underset{t\to\infty}{\lim}(\kappa_1+6\kappa_2)(t)
\,\,.
\end{equation}
The functions $\kappa_1$ and $\kappa_2$ read:
\begin{subequations}
\begin{eqnarray}\label{eq:K1}
\kappa_1(u) &=& \iiiint\limits^u \dd u_1\dd u_2\dd u_3\dd u_4
\nonumber
\\
& &\times\big[\caa\cab\cac\cad \avg{\va(u_1)\va(u_2)\va(u_3)\va(u_4)}
\nonumber
\\
& &+\caa\cab\cac\cad \avg{\xi_0(u_1)\xi_0(u_2)\xi_0(u_3)\xi_0(u_4)}
\nonumber
\\
& &+\cia\cib\cic\cid \avg{\xi(u_1)\xi(u_2)\xi(u_3)\xi(u_4)} \big]
\,\,,
\\
\kappa_2(u) &=& \iiiint\limits^u \dd u_1\dd u_2\dd u_3\dd u_4
\nonumber
\\
& &\times\big[\caa\cab\cac\cad \avg{\va(u_1)\va(u_2)} \avg{\xi_0(u_3)\xi_0(u_4)}
\nonumber
\\
& &+\caa\cab\cic\cid \avg{\va(u_1)\va(u_2)} \avg{\xi(u_3)\xi(u_4)}
\nonumber
\\
& &+\caa\cab\cic\cid \avg{\xi_0(u_1)\xi_0(u_2)} \avg{\xi(u_3)\xi(u_4)} \big]
\,\,,
\nonumber
\\
\end{eqnarray}
\end{subequations}
where $\chi_i=\chi(u-u_i)$, $\chi_{\text{\tiny A}i}=\ca(u-u_i)$, and $i\in\{1,2,3,4\}$. The non--Gaussianity of the active bursts plays a role in the first term in the bracket of Eq~\eqref{eq:K1}.
Being $\xi$ and $\xi_0$ thermal noises, their $4$--time correlation function is expressed in terms of their $2$--time correlation function as:
\begin{eqnarray}\label{eq:gauss}
\avg{\xi(\ta)\xi(\tb)\xi(\tcc)\xi(\tdd)} &=& \avg{\xi(\ta)\xi(\tb)}\avg{\xi(\tcc)\xi(\tdd)} 
\nonumber
\\
& &+ \avg{\xi(\ta)\xi(\tcc)}\avg{\xi(\tdd)\xi(\tb)}
\nonumber
\\
& &+ \avg{\xi(\ta)\xi(\tdd)}\avg{\xi(\tcc)\xi(\tb)}
\,\,,
\nonumber
\\
\end{eqnarray}
and the same property holds for the correlations of $\xi_0$. By using Eqs.~\eqref{eq:KA} and~\eqref{eq:gauss}, we finally deduce the next orders in the expression of $\avgs{x^4}$.

\subsection{Quartic optical trap}\label{app:x4}
To compute the steady state average $\avgs{x^2}$, we expand the positions $x$ and $x_0$ in terms of $\bp$ as: $x=x^{(0)}+x^{(1)}+\cO(\bp^2)$, and $x_0=x_0^{(0)}+x_0^{(1)}+\cO(\bp^2)$, where $x^{(1)}$ and $x_0^{(1)}$ are of order $\bp$. The steady state average is expressed as:
\begin{equation}
\avgs{x^2}=\avgs{\pnt{x^{(0)}}^2}+ 2\avgs{x^{(0)}x^{(1)}} +\cO(\bp^2)
\,\,.
\end{equation}
The leading order in $\bp$ equals the steady state average without quartic term in the optical trap, as we compute it in section~\ref{sec:spring}. Thus, we write the work associated with the quasistatic protocol as: $\Wqsq=\Wqsh + W_\text{\tiny P} +\cO(\bp^2)$, where $W_\text{\tiny P}=\int\dd\kp\avgs{x^{(0)}x^{(1)}}$. The positions $x^{(0)}$ and $x_0^{(0)}$ follow the dynamics in Eq.~\eqref{eq:x2}, so that the expression of $x^{(0)}$ is given by Eq.~\eqref{eq:x0}.
The positions $x^{(1)}$ and $x_0^{(1)}$ follow the coupled set of equations:
\begin{subequations}
\begin{eqnarray}
\f{\dd x^{(1)}}{\dd t} &=& -\f{1}{\td}\pnt{x^{(1)}-x_0^{(1)}} - \f{1}{\tpe} x^{(1)} - \f{\bp}{\g} \pnt{x^{(0)}}^3
\,\,,
\nonumber
\\
\\ 
\f{\dd x_0^{(1)}}{\dd t} &=& -\f{\e}{\td}\pnt{x_0^{(1)}-x^{(1)}}
\,\,,
\end{eqnarray}
\end{subequations}
from which we deduce:
\begin{equation}
x^{(1)}(t) = - \bp\int\limits^t \dd t' \chi(t-t') \pnt{x^{(0)}}^3 (t')
\,\,.
\end{equation}
We split the correlation function in the definition of $W_\text{\tiny P}$ in three contributions:
\begin{equation}
\avgs{x^{(0)}x^{(1)}}=-\g\bp\underset{t\to\infty}{\lim}\pnt{C_1+C_2+C_3}(t)
\,\,,
\end{equation}
where the functions $C_1$, $C_2$,and $C_3$ read:
\begin{subequations}
\begin{eqnarray}
C_1(t) &=& \iint\limits^t \dd u \dd s \chi(t-u) \ca(t-s) \avg{\va(s)\pnt{x^{(0)}}^3 (u)}
\,\,,
\nonumber
\\
\\
C_2(t) &=& \iint\limits^t \dd u \dd s \chi(t-u)\ca(t-s)
\nonumber
\\
& &\times\sqrt{\e\DT}\avg{\xi_0(s)\pnt{x^{(0)}}^3 (u)}
\,\,,
\\
C_3(t) &=& \iint\limits^t \dd u \dd s \chi(t-u)\chi(t-s)
\nonumber
\\
& &\times  \sqrt{\DT}\avg{\xi(s)\pnt{x^{(0)}}^3 (u)}
\,\,.
\end{eqnarray}
\end{subequations}
By using Eq.~\eqref{eq:x0}, we deduce:
\begin{subequations}
\begin{eqnarray}\label{eq:C1}
C_1(t) &=& \g^3\iint\limits^t \dd u \dd s \iiint\limits^u \dd u_1\dd u_2\dd u_3  \chi(t-u)\ca(t-s)
\nonumber
\\
& &\times\big[\caa\cab\cac \avg{\va(s)\va(u_1)\va(u_2)\va(u_3)}
\nonumber
\\
& &+3\e\DT\caa\cab\cac \avg{\va(s)\va(u_1)}\avg{\xi_0(u_2)\xi_0(u_3)} 
\nonumber
\\
& &+3\DT\caa\cib\cic \avg{\va(s)\va(u_1)}\avg{\xi(u_2)\xi(u_3)}\big]
\,\,,
\nonumber
\\
\\
C_2(t) &=& \g^3\iint\limits^t \dd u \dd s \iiint\limits^u \dd u_1\dd u_2\dd u_3  \chi(t-u)\ca(t-s)
\nonumber
\\
& &\times\big[\pnt{\e\DT}^2\caa\cab\cac \avg{\xi_0(s)\xi_0(u_1)\xi_0(u_2)\xi_0(u_3)}
\nonumber
\\
& &+3\e\DT\caa\cab\cac \avg{\xi_0(s)\xi_0(u_1)}\avg{\va(u_2)\va(u_3)}
\nonumber
\\
& &+3\e\DT^2\caa\cib\cic \avg{\xi_0(s)\xi_0(u_1)}\avg{\xi(u_2)\xi(u_3)} \big]
\,\,,
\nonumber
\\
\\
C_3(t) &=& \g^3\iint\limits^t \dd u \dd s \iiint\limits^u \dd u_1\dd u_2\dd u_3  \chi(t-u)\chi(t-s)
\nonumber
\\
& &\times\big[\DT^2\cia\cib\cic \avg{\xi(s)\xi(u_1)\xi(u_2)\xi(u_3)}
\nonumber
\\
& &+3\DT\cia\cab\cac \avg{\xi(s)\xi(u_1)}\avg{\va(u_2)\va(u_3)}
\nonumber
\\
& &+3\e\DT^2\cia\cab\cic \avg{\xi(s)\xi(u_1)}\avg{\xi_0(u_2)\xi_0(u_3)} \big]
\,\,.
\nonumber
\\
\end{eqnarray}
\end{subequations}
The non--Gaussianity of the active bursts plays a role in the first term in the bracket of Eq~\eqref{eq:C1}. From Eqs.~\eqref{eq:KA} and~\eqref{eq:gauss}, we compute the three contributions of $\avgs{x^{(0)}x^{(1)}}$, and we deduce the expression of this steady state average to leading order in $\e$:
\begin{eqnarray}
\avgs{x^{(0)}x^{(1)}} &=& -\pnt{\f{\kb\Ta}{\e}}^2 \f{\bp k^2\tau}{\kp(k+\kp)^3(\kp\tau+k(\tau+\td))}
\nonumber
\\
& &\times\brt{ 2+\pnt{\f{k}{\kp}}^2\f{5\tau+2\td}{2\tau} +\f{k}{\kp}\f{9\tau+4\td}{2\tau} }
\,\,.
\nonumber
\\
\end{eqnarray}
Finally, the expression of $\EQa$ is given by the primitive of the above formula with respect to $\kp$, thus being defined up to a constant.

\bibliographystyle{apsrev4-1}
\bibliography{energetics}

\end{document}